\documentclass{emulateapj}

\newcommand{\sqdeg}{deg$^2$}

\newcommand{\fuvband}{1350-1750\AA}
\newcommand{\nuvband}{1750-2750\AA}

\newcommand{\totband}{1350-2750\AA}
\newcommand{\galexzrange}{$0<z<2$}
\newcommand{\aislimit}{m$_{AB}\simeq 20.5$}
\newcommand{\mislimit}{m$_{AB}\simeq 23$}
\newcommand{\dislimit}{m$_{AB}\simeq 25$}

\slugcomment{Accepted for publication in Ap. J. Letters GALEX Special Issue}

\shorttitle{The Galaxy Evolution Explorer}
\shortauthors{Martin et al.}

\received{2004 June 9}
\begin{document}


\title{The Galaxy Evolution Explorer \\ A Space Ultraviolet Survey Mission}


\author{
D. Christopher Martin\altaffilmark{1},
James Fanson\altaffilmark{10},
David Schiminovich\altaffilmark{1},
Patrick Morrissey,\altaffilmark{1},
Peter G. Friedman\altaffilmark{1},
Tom A. Barlow\altaffilmark{1},
Tim Conrow\altaffilmark{1}, 
Robert Grange\altaffilmark{3},
Patrick N. Jelinsky\altaffilmark{5},
Bruno Milliard\altaffilmark{3},
Oswald H. W. Siegmund\altaffilmark{5},
Luciana Bianchi\altaffilmark{4},
Yong-Ik Byun\altaffilmark{2}, 
Jose Donas\altaffilmark{3},
Karl Forster\altaffilmark{1},
Timothy M. Heckman\altaffilmark{4},
Young-Wook Lee\altaffilmark{2},
Barry F. Madore\altaffilmark{6,7},
Roger F. Malina\altaffilmark{3},
Susan G. Neff\altaffilmark{8},
R. Michael Rich\altaffilmark{9},
Todd Small\altaffilmark{1},
Frank Surber\altaffilmark{10},
Alex S. Szalay\altaffilmark{4}, 
Barry Welsh\altaffilmark{5} and
Ted K. Wyder\altaffilmark{1}}

\altaffiltext{1}{California Institute of Technology, MS 405-47, 1200 East
California Boulevard, Pasadena, CA 91125}
\altaffiltext{2}{Center for Space Astrophysics, Yonsei University, Seoul
120-749, Korea}
\altaffiltext{3}{Laboratoire d'Astrophysique de Marseille, BP 8, Traverse
du Siphon, 13376 Marseille Cedex 12, France}
\altaffiltext{4}{Department of Physics and Astronomy, The Johns Hopkins
University, Homewood Campus, Baltimore, MD 21218}
\altaffiltext{5}{Space Sciences Laboratory, University of California at
Berkeley, 7 Gauss Way, Berkeley, CA 94720}
\altaffiltext{6}{Observatories of the Carnegie Institution of Washington,
813 Santa Barbara St., Pasadena, CA 91101}
\altaffiltext{7}{NASA/IPAC Extragalactic Database, California Institute
of Technology, Mail Code 100-22, 770 S. Wilson Ave., Pasadena, CA 91125}
\altaffiltext{8}{Laboratory for Astronomy and Solar Physics, NASA Goddard
Space Flight Center, Greenbelt, MD 20771}
\altaffiltext{9}{Department of Physics and Astronomy, University of
California, Los Angeles, CA 90095}
\altaffiltext{10}{Jet Propulsion Laboratory, California Institute of Technology, 
4800 Oak Grove Drive, Pasadena, CA 91109}

\begin{abstract}
We give an overview of the Galaxy Evolution Explorer (GALEX), a NASA Explorer Mission launched on April 28, 2003. 
GALEX is performing the first space UV sky-survey, including imaging and grism surveys in 
two bands (\fuvband\ and \nuvband). The surveys include an all-sky imaging survey 
(\aislimit), a medium imaging survey of 1000 \sqdeg (\mislimit), a deep imaging 
survey of 100 square degrees (\dislimit), and a nearby galaxy survey. Spectroscopic 
(slitless) grism surveys (R=100-200) are underway with various depths and sky coverage. Many targets overlap 
existing or planned surveys in other bands. We will use the measured UV properties of
local galaxies, along with corollary observations, to calibrate the relationship of UV and global star formation rate 
in local galaxies. We will apply this calibration to distant galaxies discovered in the
deep imaging and spectroscopic surveys to map the history of star formation in the universe over 
the redshift range \galexzrange, and probe the physical drivers of star formation in galaxies. The GALEX mission includes
a Guest Investigator program supporting
the wide variety of programs made possible by the first UV sky survey. 
\end{abstract}


\notetoeditor{This paper is submitted for consideration as the introductory article in the GALEX special issue of the Astrophysical Journal Letters.}

\section{Motivation for GALEX}

The Galaxy Evolution Explorer (GALEX), a NASA Small Explorer mission,
is performing the first all-sky imaging and spectroscopic surveys in the space ultraviolet (\totband). \footnote{Note that other missions
are performing or plan to perform nebular spectroscopic UV surveys of the diffuse UV sky, for which GALEX is obtaining
broad-band images with 5 arcsec resolution.}
The prime goal of GALEX is to study star formation in galaxies and its evolution with time. 
GALEX primary mission surveys, and dedicated observations for the 
Guest Investigator Program beginning October 2004, will also support a broad array of other investigations. 

\subsection{Galaxy Evolution}

Tinsley, in her seminal 1968 paper \citep{tinsley68}, demonstrated that passive
stellar evolution and an evolving stellar birthrate required to match the properties
of nearby galaxies had a profound impact on the appearance of galaxies observed over cosmological distances.
It is now doctrine that galaxies cannot be used as standard candles for cosmological tests.
The study of distant galaxies has become an exploration of the physical processes
that assembled luminous matter in the cores of growing dark matter halos.

Tinsley observed that the diversity of galaxies we see today is, at its core, a diversity in star formation histories.
A central goal of cosmology is to measure and explain
the star formation, gas depletion, and chemical evolution history in galaxies.
Two major developments in the mid 1990's led new urgency to this goal. The first was the discovery
of a population of star-forming Lyman Break galaxies (LBGs) at $z\sim3$ \citep{steidel96}.
The second was the prediction \citep{fallpei}, based on QSO absorption line evolution,
and first evidence \citep{lilly96} for strong evolution in the total cosmic star formation rate. These 
culiminated in the now famous plot \citep{madau96} of cosmic star formation history,
which hinted that the SFR density in the universe was a ten times more vigorous and peaked
5-8 Gyrs ago.  A major goal is to delineate the cosmic star formation history
using any and all metrics. Due in part to the diversity of techniques the current Madau plot remains a weak constraint
on cosmogenic models. A parallel effort is now underway to measure the cosmic stellar mass history \citep{2003ApJ...587...25D}--
if the stellar initial mass function is universal and constant, these histories must agree.

The rest UV provides a powerful tool for
measuring and understanding star formation in galaxies at all epochs,
a fact underscored by
the discovery and study of Lyman Break Galaxies.
As emphasized by \cite{adelberger99}, even when dust extinction
is great, rest UV luminosities remain large enough 
to be detected in UV-selected surveys. The James Webb Space Telescope will extend
rest UV selection to redshifts of 5-20, perhaps the first generation of stars. Ironically,
the interpretation of high redshift galaxies in the rest UV
is most limited by the lack of large, systematic surveys of low redshift
UV galaxies serving as a benchmark.

While the initial conditions that led to structure formation in the Universe
are becoming clear \citep{boom00,bennett03},
the formation and evolution of galaxies is tied to the complex behavior of gas dissipating 
and cooling within dark matter halos and feedback from massive stars.
Fundamentally lacking in numerical simulations and semi-analytic models is a predictive physical model for star formation.

\subsection{The UV Sky -- Precursors to GALEX}

Technological
obstacles have slowed progress in mapping the UV sky to a series of important, but incremental advances. 
The Orbital Astronomical Observatory OAO-2
\citep{code70} provide the first systematic UV photometry and spectrophotometry 
of bright stars, globular clusters and nearby galaxies. The TD-1 satellite performed an all-sky
spectrophotometric survey of objects to a visual magnitude of 9-10 \citep{boksenberg73}. The Astronomical
Netherlands Satellite (ANS)  \citep{1975A&A....39..159V}  made UV observations of stars, globular clusters, 
planetary nebulae, and galaxies. The highly successful International
Ultraviolet Explorer \citep{kondo87}, the first satellite mission to use an imaging UV detector, 
obtained thousands of targeted low and high resolution spectra in the
1200-3000\AA\ band. Along with many other results,
these targeted missions provided the foundation for galaxy stellar population synthesis models in the UV.

UV survey experiments, beginning in the 1970's, used intensified film photography and relatively small telescope apertures.
Wide field UV surveys were performed aboard Skylab \citep{1975ApJ...199L.119H},
by a lunar camera erected by Apollo 16 astronauts \citep{carruthers73}, and by the Spacelab
FAUST instrument  \citep{1993ApJ...415..875B}. 
The Ultraviolet Imaging Telescope \citep{1997PASP..109..584S} obtained 
a wealth of UV images and results over two Shuttle Astro missions.
The balloon-borne FOCA  Telescope \citep{1992IAUIn...2...49M}
obtained the first far UV luminosity function
for galaxies in the local universe \citep{1998MNRAS.300..303T} and the
first rest UV anchor point for the star formation history plot.

\subsection{GALEX Goals}

The GALEX mission was therefore designed with three overarching primary science goals.
All three goals require the primary GALEX UV surveys and multiwavelength corrolary data.
GALEX surveys are therefore designed to exploit existing and planned surveys in other bands.

\begin{figure}
\epsscale{0.85}
\plotone{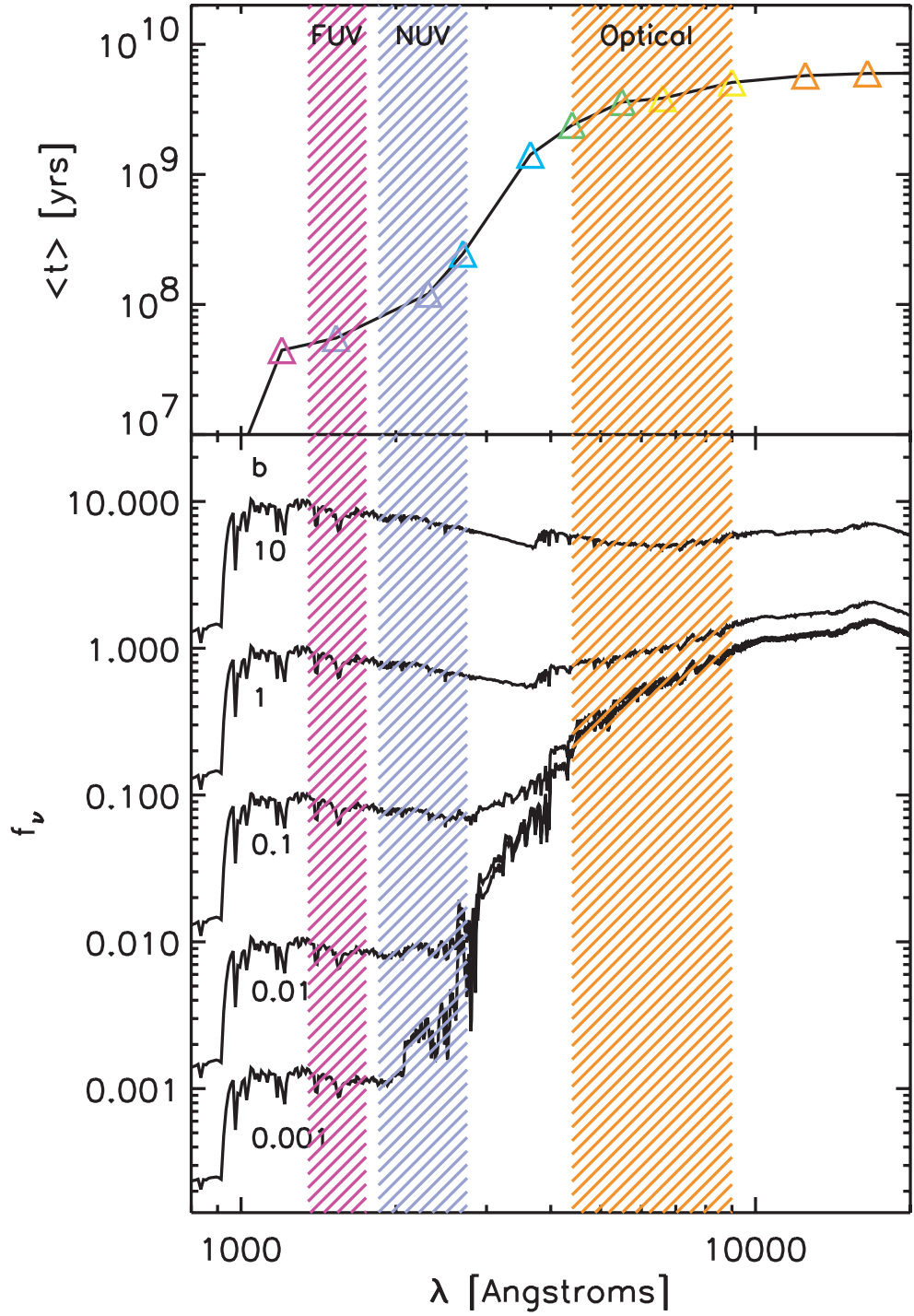}
\caption{TOP: Flux-weighted age of a Simple Stellar Population from Bruzual
and Charlot (2003), vs. wavelength. UV traces star formation over timescales of 10$^{7.5-8.5}$ yrs.
BOTTOM: Flux from old plus young stellar population, for values of $b=\dot{M}/<\dot{M}>$ (the ratio
of present to average SFR)
ranging from b=10 to b=0.001.
\label{fig_tmean_ssfr}}
\end{figure}

The first goal is to provide a calibration of UV and galaxy
star formation rate, accounting for, in order of declining impact, extinction, 
starburst history,  initial mass function and metallicity.
This calibration would be obtained over a wide range of star formation environments and
modalities, so that the relationships can be applied to galaxies at cosmic epochs where star formation may assume
a very different character. As we show in Figure \ref{fig_tmean_ssfr}, UV provides
a measure of star formation on timescales $\sim$10$^{8}$ yrs. In galaxies with smoothly varying
star formation histories, UV provides a linear measure of the current star formation rate
once the extinction problem is solved. A major GALEX objective is to determine the minimum set
of observations required to measure intrinsic extinction, in the spirit of the starburst
UV slope-infrared excess relationship of \cite{1999ApJ...521...64M}. In galaxies
with more complex star formation histories, UV probes timescales relevant to starbursts triggered by major interactions and mergers,
a core element of hierarchical structure models.

The second GALEX goal is to use the rest UV surveys to determine the cosmic star formation
history over the redshift range \galexzrange\ (the last 9 Gyrs) and 
its dependence on environment, mass, morphology, merging, and star formation modality
(notably quiescent vs. interacting/merging). This history could then be laid side by side
with ground, HST, JWST, and other optical-near IR surveys of rest UV galaxies at redshifts 1.5$<z<$20 (the first 4 Gyrs)
to yield a consistent measurement of galaxy building over the age of the universe.
A key strength of rest UV observations is the decoupling of recent and old star formation histories.
As we show in Figure \ref{fig_tmean_ssfr}, the rest UV scales with star formation rate
over a very wide range of specific star formation rates, parameterized by $b$, the ratio of present
to average star formation rates over the life of the galaxy.

Finally, the third GALEX goal is to use both large statistical samples and detailed studies of nearby galaxies,
again with abundant multiwavelength data, to inform and inspire a predictive model
of global star formation rates in diverse contexts. 

\section{Mission Overview}

\subsection{Observatory Design}

The GALEX instrument employs a novel optical design using a
50 cm diameter modified Ritchey-Chr\'etien telescope with four channels: 
Far-UV (FUV) and Near-UV (NUV) imaging, and FUV and NUV spectroscopy. The telescope has a 3-m focal length and is coated with Al-MgF$_2$.
The field of view is 1.2 degrees circular.
An optics wheel place a CaF$_2$ imaging window, a CaF$_2$ transmission grism, or an opaque position in the beam.
Spectroscopic observations are obtained at multiple grism-sky dispersion angles, also selectable, to remove spectrum overlap effects.
The FUV (\fuvband) and NUV (\nuvband)
bands are obtained simultaneously using a dichroic beam splitter that also acts as a field aberration corrector. 
The beam splitter/asphere is an ion-etched fused-silica plate with aspheric surfaces on both sides.
Beam splitting is accomplished with a dielectric multilayer on the input side, which reflects the
FUV band and transmits the NUV band.
The detector system (provided by a team led by U.C.Berkeley including Caltech, JPL, and Southwest Research) 
encorporates sealed tube microchannel plate detectors with 65 mm active area
and crossed delay-line anodes for photon event position readout.
The FUV detector is preceded by a
blue-edge filter that blocks the night-side airglow lines of OI1304, 1356, and Ly$\alpha$. The NUV detector 
is preceded by a red blocking filter/fold mirror, which produces a sharper long-wavelength cutoff than
the detector CsTe photocathode and thereby reduces the zodiacal light background and optical
contamination. 
The NUV detector has a MgF$_2$ window which includes power for field flattening,
and an opaque CsI photocathode on the microchannel plate. The NUV detector has a fused 
silica window which also includes power for field flattening,
and a semitransparent Cs$_2$Te photocathode on the window inner surface proximity focused across a
300 $\mu$m gap. The detector peak QE is 12\% (FUV) and 8\% (NUV). In orbit dark background is low, 20/60 cps (FUV/NUV) 
for diffuse background, as compared to the lowest total nightsky backgrounds of 1500/10,000 cps.
The detectors are linear up to a local (stellar) countrate of 100 (FUV), 400 (NUV) cps, which corresponds to
m$_{AB}\sim 14-15$.
The system angular resolution, which includes contributions from the optical and detector PSF, and
ex post facto aspect reconstruction, is typically 4.5/6.0 (FUV/NUV) arseconds (FWHM), and varies
by $\sim$20\% over the field of view due to variations in the detector resolution.
The grism, fabricated by Jobin-Yvon in Paris, is a ruled CaF$_2$ prism with a small curvature on the input side.
In order to provide simultaneous coverage of the \totband\ range, the
grism is blazed in 1st order for the NUV band and in 2nd order for the FUV band, and obtains
peak absolute efficiencies of 80\% and 60\% respectively.	
Spectral resolution is 200/100 (FUV 2nd order/NUV 1st order).
On orbit angular resolution and instrument throughput are as expected from ground calibration.
A more complete description of the instrument and satellite can be found in \cite{martin03}.

\subsection{Mission Operations}

GALEX was launched by a Pegasus-XL vehicle on April 28, 2003 into a 29 degree inclination, 690 km circular, 98.6 minute period orbit.
GALEX began nominal operations on August 2. A summary of on-orbit performance is given in Table \ref{tab_summary}, and more details on the on-orbit performance
are given in the following paper by \cite{morrissey04}.  The eight surveys listed in Table \ref{tab_surveys} are being performed concurrently for 
the first 38 months. The mission design is simple. Science data is obtained only on the night side. On the day side,
the satellite is in solar panel-sun orientation. As the satellite enters twilight, it slews to one of the survey 
targets. The imaging window or grism is selected, and the detector high voltage ramped from idle.
All observations are performed in a pointed mode with an arcminute spiral dither
to average non-uniformities and to prevent detector fatigue by bright stars.  
Individual photon events, time-tagged to 1-5 msec accuracy, are stored on the spacecraft solid-state tape 
recorder along with housekeeping data.  At the end of each orbital night, detector high voltages are ramped back to idle levels to 
protect detectors from damage and the spacecraft returns to solar array pointed attitude. Up to four times per 24 hour day the 
solid state recorder is dumped via the X-band transmitter to ground stations in Hawaii or Perth, Australia.  
 
The GALEX data analysis pipeline operated at the Caltech Science Operations Center receives the 
time-tagged photon lists, instrument/spacecraft housekeeping and satellite aspect information.
From these data sets, the pipeline reconstructs the 
aspect vs. time and generates images, spectra and source catalogs.  The first pipeline module 
corrects the photon positions for detector and optical distortions and calculates an optimal 
aspect solution based on the time-tagged photon data and star catalogs. 
A photometric module accumulates the photons into count, intensity  
and effective area maps and extracts sources using the well known  
Sextractor program, augmented by significant  
pre-processing of the input images, as well as post-processing the  
resulting source catalog.
A spectroscopic module uses GALEX image source catalog inputs to extract spectra of these 
sources from the multiple slitless grism observations.

\subsection{Survey Design}

Night sky backgrounds in the GALEX bands are low: the FUV[NUV] 
is dominated by diffuse galactic light[zodiacal light], with typical levels 27.5[26.5] mag arcsec$^{-2}$ corresponding 
to 3[30] photon/PSF in one eclipse. Surveys become background limited at m$_{AB}\sim$23.5. 
Targets are constrained by the current bright star limit for the 
detectors (5 kcps) which makes 50\% of the sky inaccessible. With the NUV detector off, 72\% of the sky is accessible.
This limit will be raised to 50 kcps shortly after ground verification tests. 
With this new limit, 95\% of the sky can be
surveyed in FUV-only mode, and 87\% with both detectors operating. 

{\it All-sky Imaging Survey (AIS).} The goal of the AIS is to survey the entire sky subject 
to a sensitivity of \aislimit, comparable to the POSS II 
(m$_{AB}$=21) and SDSS spectroscopic (m$_{AB}$=17.6) limits. Several hundred to 1000 objects 
are in each 1 \sqdeg~field. The AIS is performed in roughly ten 100-second pointed exposures per 
eclipse ($\sim$10 \sqdeg~per eclipse).  

{\it Medium Imaging Survey (MIS).} The MIS covers 1000 \sqdeg, with extensive overlap 
of the Sloan Digital Sky Survey.  MIS exposures are a single eclipse, typically 1500 seconds, 
with sensivity \mislimit, net several thousand objects, and are well-matched to SDSS photometric limits.

{\it Deep Imaging Survey (DIS).} The DIS consists of 20 orbit (30 ksec, \dislimit) exposures, over 80 \sqdeg, 
located in regions where major multiwavelength efforts are already underway.  
DIS regions have low extinction, low zodaical and diffuse galactic backgrounds, contiguous pointings of 10 \sqdeg~ 
to obtain large cosmic volumes, 
and minimal bright stars. A Ultra DIS of 200 ksec, m$_{AB}\sim 26$ is also in progress in  four fields.



\begin{figure*}
\epsscale{1.0}
\plotone{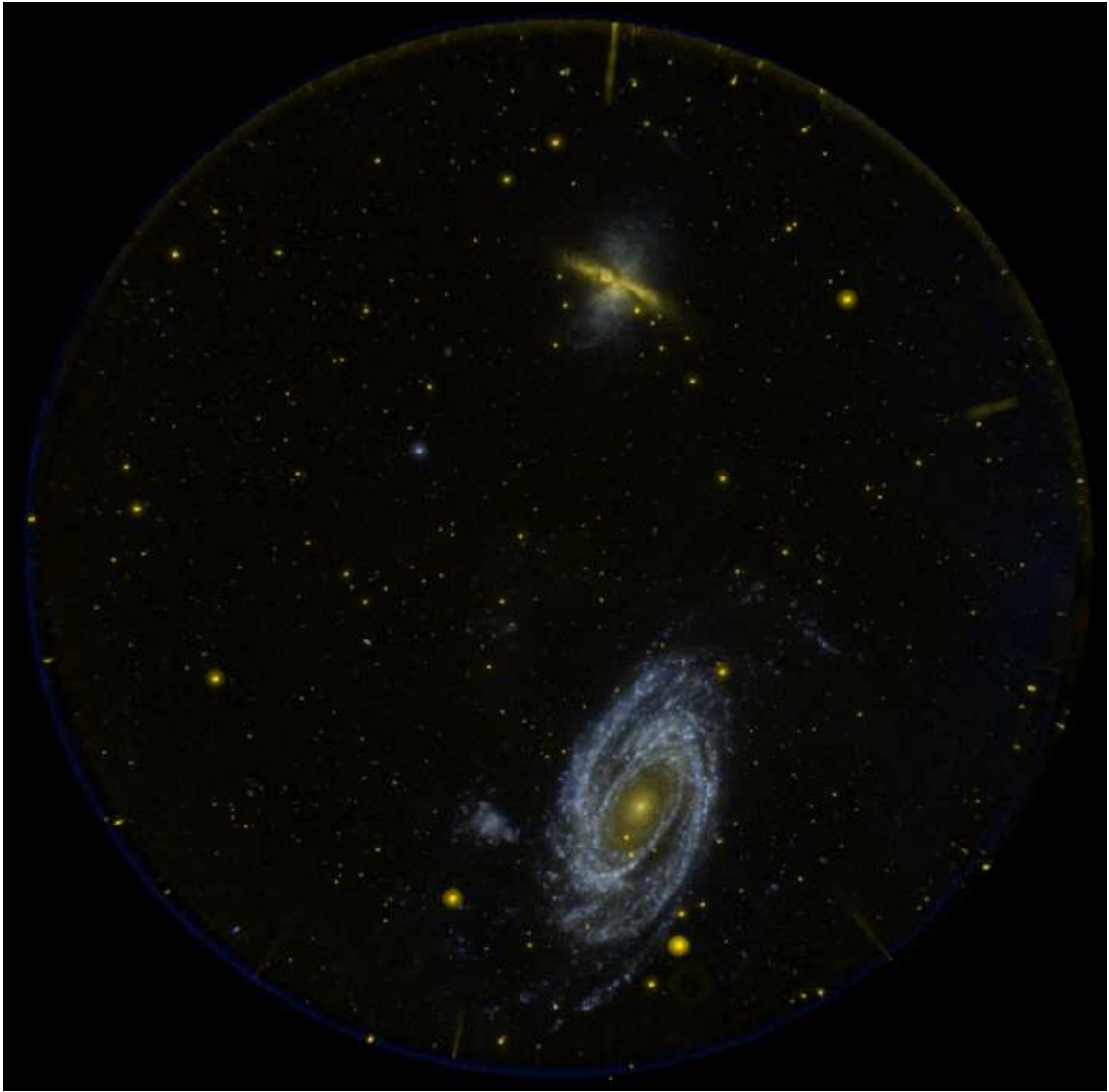}
\caption{In Figure \ref{fig_m81m82}, we show the GALEX NGS observation of the M81-M82 system.
This figure illustrates how a single GALEX image may be used to study
many different aspects of galaxy evolution: spiral structure, stellar populations,
extinction (M81), interaction-induced starbursts and resulting galactic outflows (M82);
star formation in dwarf galaxies (Ho IX), between galaxies and in tidal streams; the UV background;
and large scale structure. In the color table, red-green (gold) is used for NUV, and blue for FUV. \label{fig_m81m82}}
\end{figure*}

{\it Nearby Galaxies Survey (NGS).} The NGS targets well-resolved nearby galaxies for 1-2 eclipses.  
Surface brightness limits are m$_{AB}\sim$27.5 arcsec$^{-2}$, or a star formation rate of 
10$^{-3}$ M$_\odot$ yr$^{-1}$ kpc$^{-2}$.  The 200 targets are a diverese selection
of galaxy types and environments, and include most galaxies from the 
Spitzer IR Nearby Galaxy Survey (SINGS).
Figure \ref{fig_m81m82} shows the NGS observation of the M81/M82 system. 

{\it Spectroscopic Surveys.}  The suite of spectroscopic surveys includes (1) the Wide-field Spectroscopic Survey (WSS), which covers the full 80 \sqdeg~DIS footprint with comparable exposure time (30 ksec), and reaches m$_{AB}\sim 20$ for  S/N$\sim$10 spectra; (2) the Medium Spectroscopic Survey (MSS), which covers the high priority central field in each DIS survey region (total 8 \sqdeg) to m$_{AB}$=21.5-23, using 300 ksec exposures; and (3) Deep Spectroscopic Survey (DSS) covering 2 \sqdeg~with 1000 eclipses, to a depth of m$_{AB}$=23-24. 

\section{Early Results}

In this dedicated Astrophysical Journal Letters issue, we describe a selection of early results
obtained from the GALEX surveys. In this introductory letter, we give examples of data from
the various surveys and highlight some of the early results described in more detail in this issue.

\subsection{Star Formation in Diverse Contexts}

GALEX observations demonstrate that rest UV emission traces star formation in a wide variety of
contexts, environments and modalities. In nearby quiescent disk galaxies (M101, M51, M33)
GALEX observations  provide the ages, luminosities, masses, and extinction of star formation
complexes. These show that the cluster age distribution is consistent with a constant star formation rate
over the last 10$^9$ Myrs \citep{bianchi04}.

GALEX has discovered extended UV emission far outside (2-4 times) the optical disk of a number of nearby
spiral galaxies, including M83 and NGC628. Star formation appears to proceed at gas surface
densities below the typical galaxian threshold, and the complexes
formed appear to have lower luminosity and mass, and younger ages than those in the
inner disk \citep{thilker04}. 

In the Antennae merger system, recent star formation has occured 
in the disk, tails and Tidal Dwarf Galaxy on timescales less than
the 300 Myr dynamical time, implying that star formation is 
triggered in the tidal streams after gas leaves the
galaxy \citep{hibbard04}.  GALEX has detected UV emission in extended tidal tails of a number of other
interacting galaxies \citep{neff04a,xu04b} showing ages less than interaction times, suggesting that
interaction-induced star formation in tidal gas streams may be a common phenomenon.
Star formation triggered in an HI cloud 50 kpc from Centaurus A \citep{neff04b}
is easily detected in the UV.  All of these phenomena are likely to
be more common at high redshift.

\subsection{Star Formation History}

Large area, unbiased, multiwavelength surveys support a robust statistical study of the
fundamental properties of galaxies. GALEX provides an unprecedented measurement of
star formation rates. When combined with rest optical/near IR, the key parameter of star formation
rate per unit stellar mass (specific star formation rate), which is closely related to
the current star formation rate with respect to the average (the $b$ parameter), can be determined over a wide dynamic range.

GALEX combined with the Sloan Digital Sky Survey forms a powerful dataset. GALEX-SDSS colors
provide preliminary source classification and characterization \citep{seibert04a}, separating
main sequence, post-main-sequence, and binary stars, QSOs, and galaxies. GALEX/SDSS photometry
provide excellent measurement of the ratio of current to average star formation rates (the $b$-parameter)
and important constraints on starburst history in the local universe \citep{salim04}.

In the local universe UV luminosities follow a Schechter function with $L_*=2\times 10^9$L$_\odot$ \citep{wyder04},
with measureably different parameters for red and blue subsamples and evidence for evolution \citep{treyer04}.
The luminosity functions and densities are significantly fainter than those
obtained by FOCA.  GALEX on-orbit calibration shows excellent agreement with the ground calibration
\citep{morrissey04}, so we attribute this discrepency to an overestimate of fluxes at faint magnitudes by FOCA.
GALEX number count distributions fall below those of FOCA, show some evidence for evolution,
and show some flattening at faint magnitudes due to a combination of blending, confusion, and
faint wing overlap \citep{xu04a}. 

GALEX DIS data, combined with the VLT/VIMOS redshift survey, has provided the first
GALEX measurements of the evolution of the UV luminosity
function and density \citep{arnouts04}. Significant evolution is found, consistent with $\sim(1+z)^2$.
When corrected for extinction using the \cite{kong04} prescription, this provides the
first GALEX measurements of cosmic SFR history over $0<z<1.2$ \citep{schiminovich04}. When combined
with consistent analysis of deep optical data from HDF-N \citep{arnouts04,schiminovich04}, the results
suggest a monotonic decline in the star formation rate density since z=3, rather than a peak at $z\sim 1-1.5$.

The GALEX AIS/MIS-SDSS DR1 matched datasets have yielded the discovery of Luminous UV Galaxies (LUGs),
with UV luminosities and properties comparable to distant Lyman Break Galaxies, in the local universe \citep{heckman04}.
LUGs have 500 times lower volume density than LBGs, but their density rises sharply toward
higher redshift \citep{schiminovich04}. LUGs may provide an excellent opportunity
to study low redshift analogs to massive star-forming galaxies at high redshift.

\subsection{Dust Extinction}

UV extinction by dust remains the principle obstacle in converting UV luminosity 
directly to SFR.  In individual galaxies
(M83 and M101), we show that extinction is a strongly declining function of radius
and may contain a significant diffuse, interarm component \citep{popescu04,boissier04}.

With IRAS data, we have begun 
to characterize the properties of FUV and Far-IR selected galaxy samples in the local 
universe. As expected, flux-limited UV and Far-IR selected samples yield different projections 
of the bivariate UV/Far-IR luminosity function and distinct UV/Far-IR ratios \citep{buat04}.  However, virtually 
all star-forming galaxies detected in local Far-IR-selected samples are also detected by GALEX. 
The SFR luminosity function places a fundamental constraint 
on cosmological models.   The bivariate UV/FIR luminosity function, obtained from combined UV and FIR selected samples,
shows a bimodal behavior: $L_{FUV}$ tracks $L_{FIR}$
for $L_{TOT}<3 \times 10^9$ L$_\odot$, and $L_{FUV}$ saturates for higher total luminosities.
The total SFR function is log-normal function over four decades of luminosity \citep{martin04b}.

Starburst galaxies in the local universe display a well-known relationship between UV slope ($\beta$) and 
Far-IR-to-UV luminosity ratio (IRX).  Early results from GALEX indicate that 1) normal and 
quiescent star forming galaxies fall below the canonical IRX-$\beta$  relation for 
starbursts, perhaps due to starburst age \citep{kong04}; and 2) the difference between starbursts and ULIRGs 
within IRX-$\beta$ space is not as distinct as previously suggested, which is consistent with 
our observation that galaxies span a continuum in the bivariate UV/Far-IR luminosity function.  

\subsection{UV from Stars, Gas, Dust, and AGN}

In addition to recent star formation, UV traces dust through scattering and absorption, gas by emission
lines and two-photon continuum, hot evolved stars and late-type stellar chromospheres, degenerate
binaries, and QSOs.  While some 15-20\% of ``normal'' elliptical galaxies show evidence for residual star formation
at the 1-2\% level \citep{yi04}, quiescent ellipticals (notably in rich clusters) exhibit
the well known UV excess but without the previously claimed correlation with metallicity \citep{rich04}.
The M82 outflow, prominent in Figure \ref{fig_m81m82}, is injecting abundant dust into the IGM, which
reflects UV from the obscured staburst \citep{hoopes04}. In our own galaxy, the remarkable Criss-Cross nebula
shines brightly in two-photon continuum from a moderate velocity shock \citep{seibert04c}.
As the AIS progresses, the number of QSOs suitable for measurements of HeII reionization are increasing \citep{tytler04}.


\section{GALEX Data Legacy}

GALEX Early Release Data was made available in December, 2003. The first major
data release, GDR1, will occur on October 1, 2004, coincident with the start of the
GALEX Guest Investigator Program. All GALEX data are served by the STScI
MAST Archive. While we cannot predict
the applications to which the GALEX data will be applied in the future, we can anticipate
that the impacts of the first comprehensive UV Sky Survey will be broad and lasting.

\acknowledgments

GALEX (Galaxy Evolution Explorer) is a NASA Small Explorer, launched in April 2003.
We gratefully acknowledge NASA's support for construction, operation,
and science analysis for the GALEX mission,
developed in corporation with the Centre National d'Etudes Spatiales
of France and the Korean Ministry of 
Science and Technology. The grism, imaging window, and uncoated aspheric corrector were supplied by France.
We acknowledge the dedicated
team of engineers, technicians, and administrative staff from JPL/Caltech, 
Orbital Sciences Corporation, University
of California, Berkeley, Laboratory Astrophysique Marseille, 
and the other institutions who made this mission possible.


\clearpage
\clearpage

\clearpage

\begin{table}
\caption{GALEX On-Orbit Performance\label{tab_summary}}
\begin{tabular}{ll}\tableline
Effective Area          &       20-50 cm$^2$            \\
Angular resolution      &       4.5-6\arcsec FWHM               \\
Spectral Resolution     &               100-250         \\
Field of View           &       1.2 degrees                     \\
Bands   (simultaneous)          &       FUV \fuvband; NUV \nuvband      \\
Sensitivity                     &       100 s    20.5  [AIS]    \\
(AB mag)                        &       1 ks     23.5  [MIS/NGS]\\
                                &       30 ks    25.5  [DIS]    \\
Astrometry                      &       1 arsec (rms) \\
Observations            &       Nightime--1 eclipse=1000-2000 s \\
Mission Length          &       Baseline 38 months, 13 months to date   \\ \tableline
\end{tabular}
\end{table}
\clearpage

\begin{table}
\caption{Survey Summary\label{tab_surveys}}
\begin{tabular}{lcccccc}\tableline
        Survey&         Area&           Expos&  m$_{AB}$&       \#Gals& Volume& $<z>$ \\
        &               [deg$^2$]&              [ksec]& &       (est.)& [Gpc$^3$]& \\ \tableline
        All-Sky Imag. (AIS)     &40,000         &0.1    &20.5           &10$^7$ &1& 0.1 \\ \tableline
        Medium Imag. (MIS)      &1000   &1.5    &23     &$10^{6.5}$&    ~1&     0.6                     \\ \tableline
        Deep Imaging (DIS)&     80&             30&     25&     107&    1.0&    0.85 \\ \tableline
        Ultra-Deep Imag.(UDIS)&         1&      200&    26&     $10^{5.5}$&     0.05&   0.9\\ \tableline
        Nearby Galaxies (NGS)   &---    &0.5            &27.5\tablenotemark{a}& 200&    ---&    --              \\ \tableline
        Wide Spectro.(WSS)&     80&             30&     20&     10$^{4-5}$&     0.03&   0.15\\ \tableline
        Medium Spectro. (MSS)&  8&              300     &21.5\tablenotemark{b}& 10$^{4}$&       0.03    &0.3    \\
                                        &       &               &23.3\tablenotemark{c}& 10$^{5}$&       0.03    &0.5    \\ \tableline
        Deep Spectro. (DSS)&    2&              2000&   22.5\tablenotemark{b}&  10$^{4}$&       0.05    &0.5            \\
                                &       &               2000&   24.3\tablenotemark{c}&  10$^{5}$&               &0.9            \\ \tableline
\end{tabular}
\tablenotetext{a}{mag. sq. arcssec $^b$ R=100; $^c$R=20}

\end{table}

\end{document}